\documentclass[twocolumn]{aastex631}

\usepackage{amsmath}

\shorttitle{Polarization observations of the blazar Mrk 421}
\shortauthors{Athira M. Bharathan et al.}

\graphicspath{{./}{figures/}}

\begin{document}

\title{Simultaneous X-ray and optical polarization observations of the blazar Mrk 421}

\correspondingauthor{Athira M. Bharathan}
\email{athira.bharathan@res.christuniversity.in}

\correspondingauthor{Blesson Mathew}
\email{blesson.mathew@christuniversity.in}

\author{Athira M. Bharathan}
\affiliation{Department of Physics and Electronics, CHRIST (Deemed to be University), Bangalore, India} 

\author{C. S. Stalin}
\affiliation{Indian Institute of Astrophysics, Block II, Koramangala, Bangalore 560 034, India} 

\author{S. Sahayanathan}
\affiliation{Astrophysical Sciences Division, Bhabha Atomic Research Centre, Mumbai-400085, India}
\affiliation{Homi Bhabha National Institute, Mumbai-400094, India}

\author{Kiran Wani}
\affiliation{Aryabhatta Research Institute for Observational Sciences, Manora Peak, Nainital, India}
\affiliation{School of Physical Sciences, SRTM University, Nanded, 431 606, India}

\author{Amit Kumar Mandal}
\affiliation{Astronomy Program, Department of Physics and Astronomy, Seoul National University, Seoul 151-742, Republic of Korea}
\affiliation{Center for Theoretical Physics, Polish Academy of Sciences, Al. Lotnikow´ 32/46, PL-02-668 Warsaw, Poland}

\author{Rwitika Chatterjee}
\affiliation{Space Astronomy Group, ISITE Campus, U. R. Rao Satellite Centre, Bangalore 560 037, India.}

\author{Santosh Joshi}
\affiliation{Aryabhatta Research Institute for Observational Sciences, Manora Peak, Nainital, India}

\author{Jeewan C Pandey}
\affiliation{Aryabhatta Research Institute for Observational Sciences, Manora Peak, Nainital, India}

\author{Blesson Mathew}
\affiliation{Department of Physics and Electronics, CHRIST (Deemed to be University), Bangalore, India}

\author{Vivek K. Agrawal}
\affiliation{Space Astronomy Group, ISITE Campus, U. R. Rao Satellite Centre, Bangalore 560 037, India.}

\begin{abstract}
We present near-simultaneous X-ray and optical polarization measurements in the high synchrotron peaked (HSP) blazar Mrk 421. The X-ray polarimetric observations were carried out using {\it Imaging X-ray Polarimetry Explorer} ({\it IXPE}) on 06 December 2023. During {\it IXPE} observations, we also carried out optical polarimetric observations using 104cm Sampurnanand telescope at Nainital and multi-band optical imaging observations using 2m Himalayan Chandra Telescope at Hanle. From model-independent analysis of {\it IXPE} data, we detected X-ray polarization with degree of polarization ($\Pi_X$) of 8.5$\pm$0.5\% and an electric vector position angle ($\Psi_X$) of 10.6$\pm$1.7 degrees in the 2$-$8 keV band. From optical polarimetry on 06 December 2023, in B, V, and R bands, we found values of $\Pi_B$ = 4.27$\pm$0.32\%, $\Pi_V$= 3.57$\pm$0.31\%, and $\Pi_R$= 3.13$\pm$0.25\%. The value of $\Pi_B$ is greater than that observed at longer optical wavelengths, with the degree of polarization suggesting an energy-dependent trend, gradually decreasing from higher to lower energies. This is consistent with that seen in other HSP blazars and favour a stratified emission region encompassing a shock front. The emission happening in the vicinity of the shock front will be more polarized due to the ordered magnetic field resulting from shock compression. The X-ray emission, involving high-energy electrons, originates closer to the shock front than the optical emission. The difference in the spatial extension could plausibly account for the observed variation in polarization between X-ray and optical wavelengths. This hypothesis is further supported by the broadband spectral energy distribution modeling of the X-ray and optical data.

\end{abstract}


\section{Introduction} \label{sec:intro}

Blazars are a unique category of active galactic nuclei (AGN) that are powered by accretion of matter onto supermassive black holes situated at the centre of galaxies \citep{1969Natur.223..690L,1973A&A....24..337S,1995PASP..107..803U}. Their emission spans the accessible electromagnetic spectrum from low-energy radio to high-energy $\gamma$-rays \citep{2019NewAR..8701541H}. The other observational properties of blazars include high bolometric luminosity, a two hump broadband spectral energy distribution (SED; \citealt{1998MNRAS.299..433F}), dominated by non-thermal emission, strong flux variability across the entire electromagnetic spectrum with time scale of variations ranging from minutes to hours \citep{1995ARA&A..33..163W,1997ARA&A..35..445U}, high degree of polarization in the radio and optical bands \citep{1980ARA&A..18..321A}, optical polarization variability \citep{2005A&A...442...97A} and super-luminal motion \citep{1995PASP..107..803U}. These observational characteristics of blazars are likely due to their relativistic jets oriented close to the line of sight to the observer. High synchrotron peaked (HSP) blazars, a subset of blazars, have the peak of their low-energy component occurring at frequencies greater than 10$^{15}$ Hz in their broadband SED \citep{2010ApJ...716...30A}.
Blazars have been extensively studied using various diagnostics to understand the emission processes in them. Of these, multi-wavelength linear polarization observations can serve as an efficient tool to constrain blazar emission. Such an investigation can provide the needed constraints on the particle acceleration processes in blazar jets. Polarization observations have historically been accessible at radio, infrared, and optical wavelengths; new developments in the X-ray range have enabled the detection of X-ray polarization with the launch of the \textit{Imaging X-ray Polarimetry Explorer} (\textit{IXPE}; \citealt{2022JATIS...8b6002W}) satellite, which began its scientific operations in January 2022.


The BL Lac object Mkn 421 is one of the closest (z = 0.031) and broadly examined blazars with a non-thermal spectrum spanning from radio to very high-energy $gamma$-rays \citep{2009ApJ...703..169A}. Given its status as the brightest HSP blazar at X-ray energies, Mrk 421 presents an ideal target for polarization measurements using  {\it IXPE} \citep{2023NatAs...7.1245D}. Hence, this source has been observed by {\it IXPE} multiple times \citep{2022ApJ...938L...7D, 2023NatAs...7.1245D,2024A&A...681A..12K}. Such X-ray observations can be used to probe the region of the jet where the flow is accelerated and collimated.
This paper reports the results of simultaneous X-ray and optical polarization observations of Mrk 421 on 06 December 2023.
The paper is structured as follows: Section \ref{sec:obs} covers the observations and data reduction. The analysis and results are presented in Section \ref{sec:anr}, followed by the discussion in Section \ref{sec:dis}. The final section (Section \ref{sec:sum}) provides a summary.

\section{Observations and Data Reduction} \label{sec:obs}
\subsection{X-ray polarization}

The blazar Mrk 421 has been observed multiple times using {\it IXPE} since May 2022 \citep{2022ApJ...938L...7D, 2023NatAs...7.1245D, 2023AAS...24133602E,2024AA...681A..12K}. This study focuses on a recent observation conducted on 06 December 2023 (OBSID: 02008199), with {\it IXPE}, for a net exposure time of 70751 seconds. Cleaned and calibrated level 2 data were used in our scientific analysis. Using {\tt IXPEOBSSIM} software version 30.0.0 \citep{2022SoftX..1901194B},  we performed the analysis using publicly available data. We utilized the {\tt CMAP} method in the {\tt xpbin} task to generate a counts map in sky coordinates. For source extraction, we utilized a circular region with a radius of $60''$ across all three Detector Units (DUs). For background extraction in each DU, we employed a source-free annular region with an inner radius of $100''$ and an outer radius of $300''$. Subsequently, the {\tt xpselect} task was utilized to create filtered source and background regions for polarimetric analysis.

           

\begin{figure*}
\centering
\includegraphics[scale=0.3]{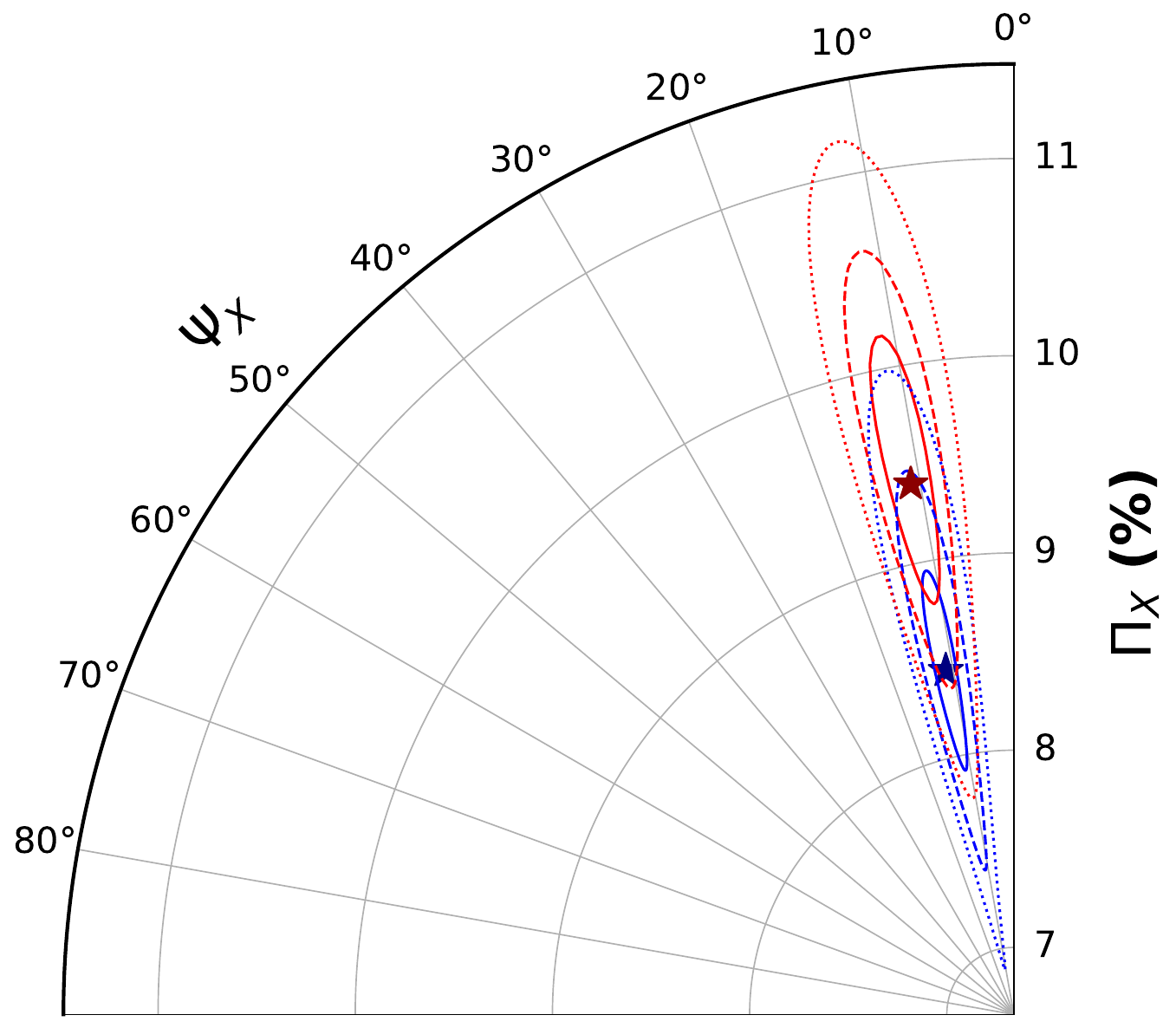}
\includegraphics[scale=0.3]{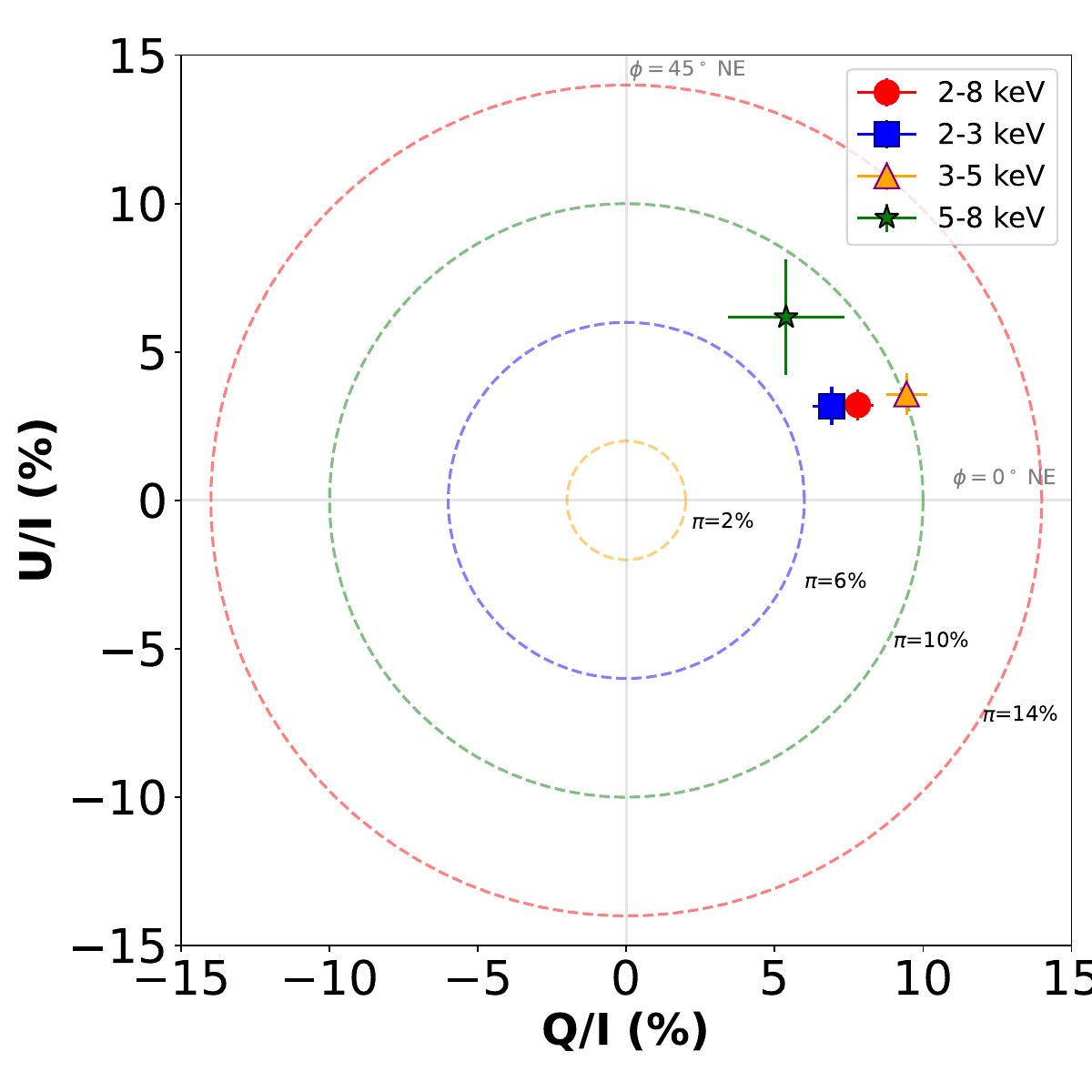}
\includegraphics[scale=0.3]{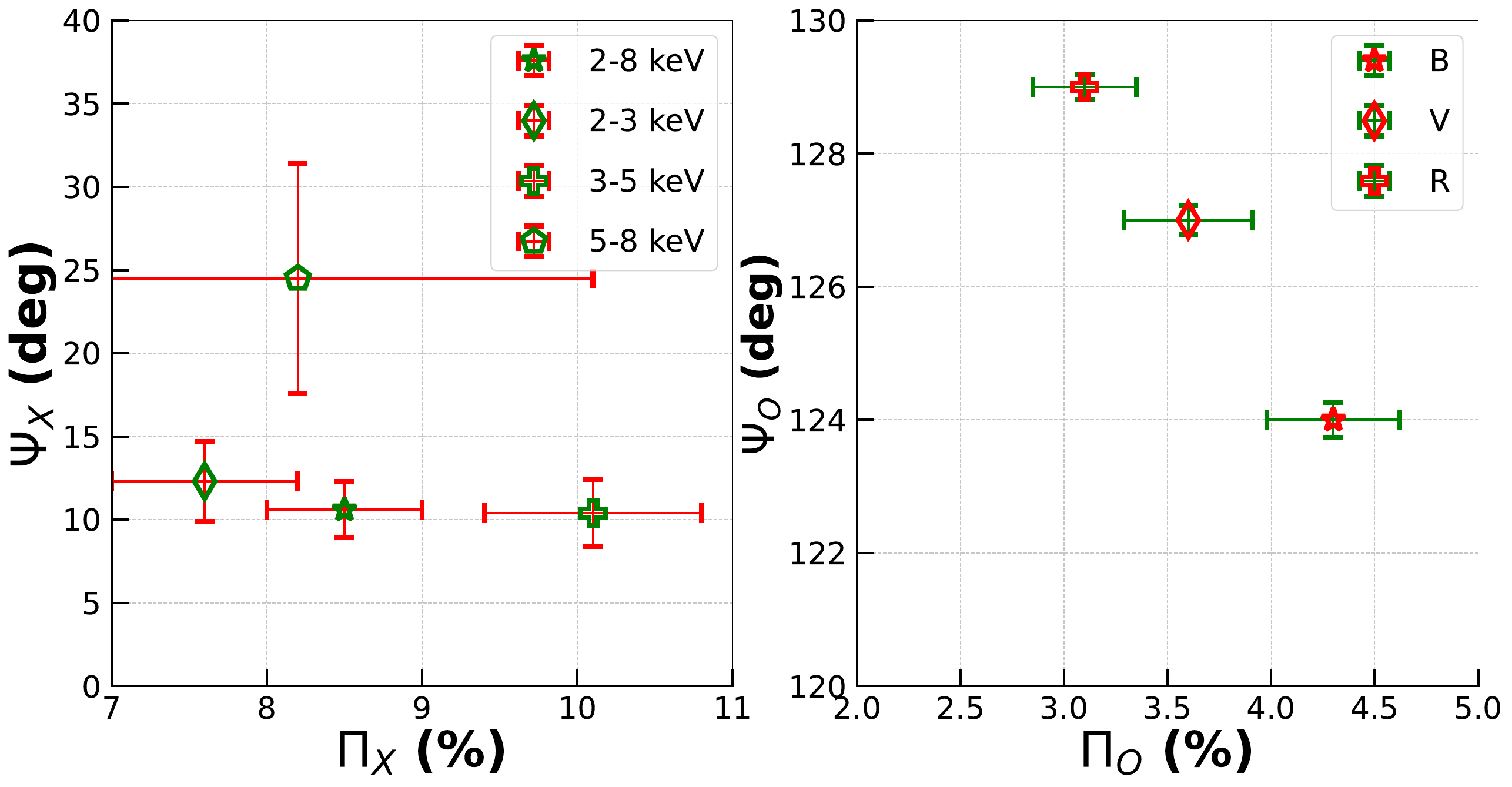}
\caption{ Results of the optical and X-ray polarization measurements on 06 December 2023. Top left: position of the 2$-$8 keV X-ray polarization in the $\Pi_{X}$ - $\Psi_{X}$ plane obtained using PCUBE (blue star) and XSPEC analysis (red star). Top right: the normalized U/I and Q/I Stokes parameter in the total 2$-$8 keV (red circle), 2$-$3 keV (blue square), 3$-$5 keV (orange triangle) and 5$-$8 keV (green star). Bottom left: the measured X-ray polarization in $\Pi_{X}$ - $\Psi_{X}$ plane . Here, star represents 2$-$8 keV, diamond represents 2$-$3 keV, plus represents 3$-$5 keV and pentagon represents 5$-$8 keV. Bottom right: The measured optical polarization ($\Pi_O$) in B-band (star), V-band (diamond) and R-band (plus).}
\label{figure-comp}
\end{figure*}

\begin{figure*}
\centering
\includegraphics[scale=0.2]{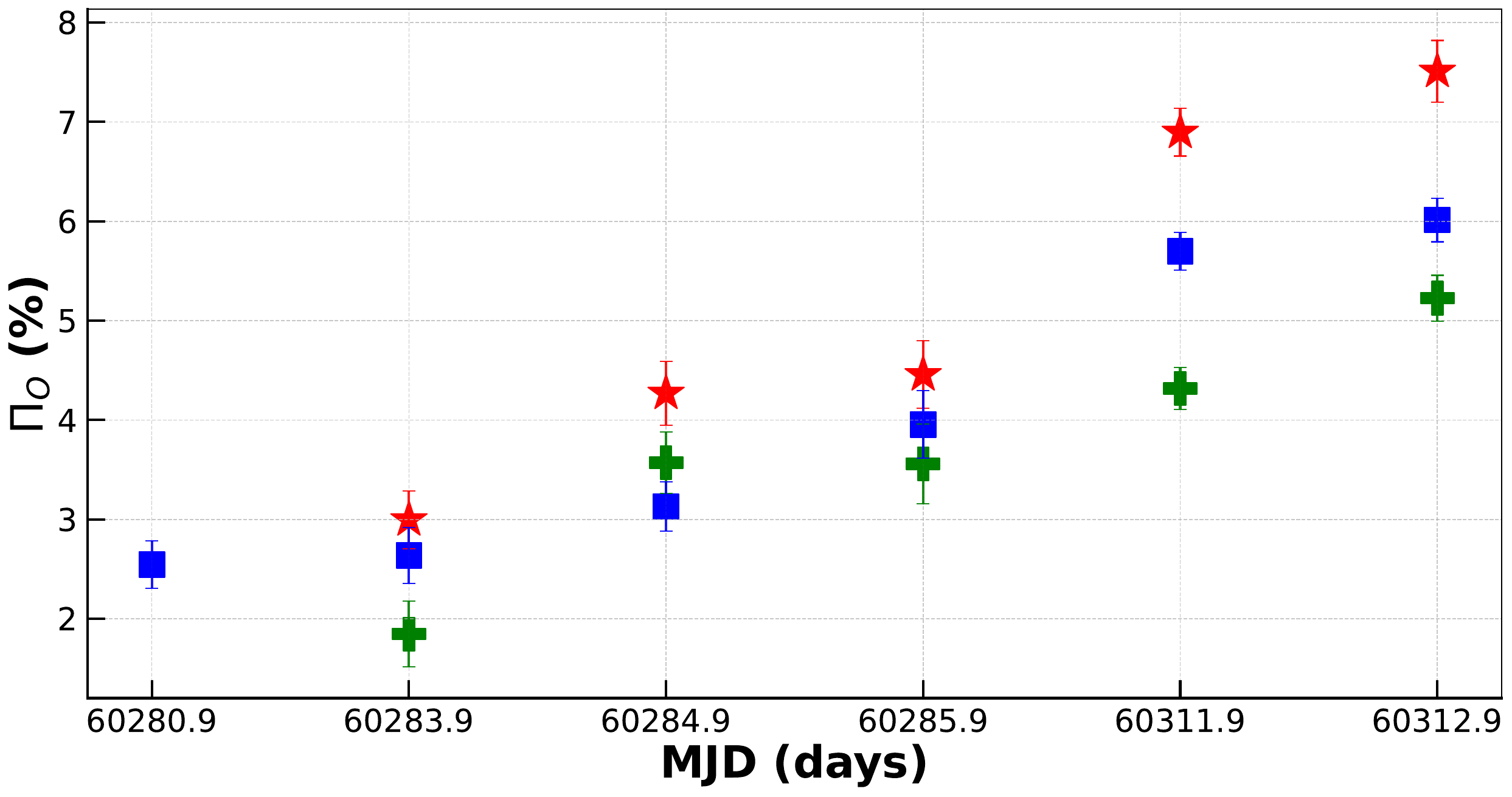}
\includegraphics[scale=0.2]{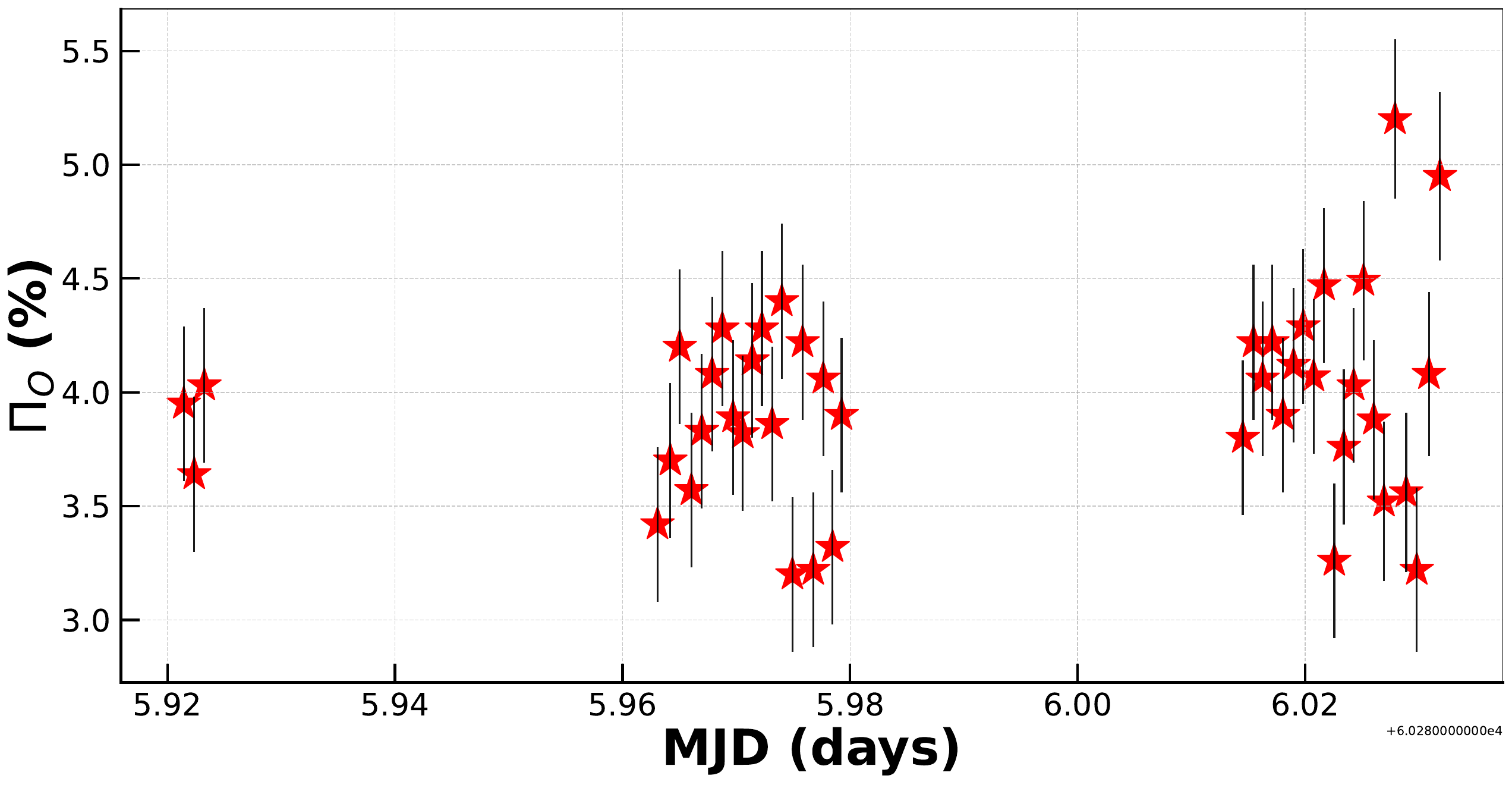}
\caption{Left panel: The measured mean optical polarization degree  in B (star), V (plus), and R (square) filters, for all the epochs of observations (see Table 5). Right panel: R-band measurements (n=42) from 07 December 2023, demonstrating significant variability in the degree of polarization over the course of a single night.}
\label{figure-ox}
\end{figure*}

\subsection{Optical polarization}
Optical polarimetric observations were carried out with the ARIES Imaging Polarimeter (AIMPOL) mounted on the 104 cm  Sampurnanand Telescope, Nainital. AIMPOL is a dual-channel polarimeter forming an extra-ordinary and ordinary image of a single object on the CCD \citep{2004BASI...32..159R, 2023JAI....1240008P}. Four different half-wave plate (HWP) rotation (i.e 0$^\circ$, 22.5$^\circ$, 45$^\circ$ \& 67.5$^\circ$) measurements are required in AIMPOL. Stokes parameters for each rotation angle were calculated as follows:
\begin{equation}
    R(\alpha)=\frac{\bigg(\frac{I_{e}(\alpha)}{I_{o}(\alpha)}-1\bigg)}{\bigg(\frac{I_{e}(\alpha)}{I_{o}(\alpha)}+1\bigg)}
\end{equation}
where I$_{e}(\alpha)$ \& I$_{o}(\alpha)$ are the intensities of the extra-ordinary and ordinary images, respectively, of the source at four different HWP rotation angles ($\alpha$). Since the response of the CCD is not the same for extra-ordinary and ordinary images of the object, the flux ratio $\frac{I_{e}(\alpha)}{I_{o}(\alpha)}$ was multiplied by a factor F and it is expressed as \citep{1998A&AS..128..369R}:
\begin{equation}
    F= \bigg[\frac{I_{o}(0^\circ)\times I_{o}(22.5^\circ)\times I_{o}(45^\circ)\times I_{o}(67.5^\circ)}{I_{e}(0^\circ)\times I_{e}(22.5^\circ)\times I_{e}(45^\circ)\times I_{e}(67.5^\circ)}\bigg]^{1/4}
\end{equation}
 Normalised four Stokes parameters corresponding to each HWP rotation angle (i.e. R(0$^\circ$)$\sim$q, R(22.5$^\circ$)$\sim$u, R(45$^\circ$)$\sim$q$_{1}$ \& R(67.5$^\circ$)$\sim$u$_{1}$) were used to calculate the degree of polarization P and the polarization position angle $\theta$. P was calculated from an average of p$_{1}$=$\sqrt{q^{2}+u^{2}}$ and p$_{2}$=$\sqrt{q_{1}^{2}+u_{1}^{2}}$, similarly $\theta$ was calculated from an average of $\theta_{1}$= 0.5$\times$tan$^{-1}(\frac{u}{q})$ \& $\theta_{2}$= 0.5$\times$tan$^{-1}(\frac{u_{1}}{q_{1}})$. Aperture photometry using the $\it{astropy }$ package-$\it{photutils}$ was used to obtain the flux values of the extra-ordinary and ordinary images of the source. We observed unpolarized and polarized standard stars during each observation night and used them respectively to correct instrumental polarization and in zero-point calibration of the position angle (PA). HD19820 and HD245310 were observed as standard polarized stars, while G191B2B and HD14069 were observed as standard unpolarized stars \citep{1992AJ....104.1563S}. We observed the source on 6 epochs between 02 December 2023 and 03 January 2024, the details of which are given in Table \ref{table-optical}.

\begin{table}
\centering
\caption{The log of optical polarization observations. Given are the dates of observations and the number of images acquired in B,V and R filters.}
\label{table-optical}
\begin{tabular}{cccc}
\hline
Date  & \multicolumn{3}{c}{Number of observations}  \\
 & B & V & R \\
\hline
02/12/2023 & 0 & 0 & 3  \\
05/12/2023 & 3 & 3 & 3  \\
06/12/2023 & 3 & 3 & 6  \\
07/12/2023 & 3 & 3 & 42  \\
02/01/2024 & 3 & 3 & 3  \\
03/01/2024 & 3 & 3 & 3  \\
\hline
\end{tabular}
\end{table}

\begin{table}
\begin{center}
\caption{Results of the X-ray polarization observations on 06 December 2023 (OBSID = 02008199). Here, MDP is the minimum detectable polarization}\label{table-xpol}
 \begin{tabular}{cccccccccccc}
  \hline\noalign{\smallskip}
& & 2$-$3 keV & 3$-$5 keV & 5$-$8 keV & 2$-$8 keV  \\
\hline\noalign{\smallskip}
$\Pi_{X}(\%)$ &  & 7.6$\pm$0.6 & 10.1$\pm$0.7 & 8.2$\pm$1.9 &  8.5$\pm$0.5  \\
$\Psi_{X}(deg)$ & & 12.3$\pm$2.4 & 10.4$\pm$2.0 & 24.5$\pm$6.9 & 10.6$\pm$1.7  \\
MDP (\%)&  & 1.95 & 2.11 & 5.94 & 1.52 \\

  \noalign{\smallskip}\hline
\end{tabular}
\end{center}
\end{table}

\subsection{Optical imaging}
Optical imaging observations were carried out in u$^{\prime}$, g$^{\prime}$, r$^{\prime}$, and
i$^{\prime}$ filters on 05 December 2023, using the 2 m Himalayan Chandra Telescope. The observed optical data were reduced using \textit{IRAF} (\textit{Image Reduction and Analysis Facility}; \citealt{1986SPIE..627..733T, 1993ASPC...52..173T}), following standard procedures including bias subtraction, dark subtraction,  flat-fielding, and cosmic rays removal. Following reduction, objects within the observed image frames were identified utilizing the \textit{daofind} task within IRAF. Subsequently, photometry was conducted on these detected objects using the \textit{phot} task in \textit{IRAF}. From the detected objects, several comparison stars from the same field of view (FoV) were selected to perform differential photometry to bring the instrumental magnitude obtained for Mrk 421 to the standard system. For this aperture photometry was performed with an aperture size three times the seeing size, typically ranging from 4.4 to 6.2 arcsec in u$^{\prime}$, g$^{\prime}$, r$^{\prime}$, and i$^{\prime}$ filters. Notably, the aperture size chosen for photometry of Mrk 421 excludes the nearby galaxy (LEDA 33453). For differential photometry the apparent magnitudes of the comparison stars were taken from the SDSS DR18 sky survey{\footnote{https://skyserver.sdss.org/dr18/VisualTools/explore/summary}} in the respective filters.

\subsection{Swift-XRT}
To generate the X-ray spectral data points needed for constructing the SED, we used observations from the X-Ray Telescope (XRT; \citealt{2005SSRv..120..165B}) onboard the Neil Gehrels Swift Observatory\footnote{https://heasarc.gsfc.nasa.gov/cgi-bin/W3Browse/swift.pl}. These observations were conducted simultaneously with \textit{IXPE} on 06 December 2023 (OBSID: 00097201001). We processed the XRT data using the {\it xrtpipeline}\footnote{https://www.swift.ac.uk/analysis/xrt/xrtpipeline.php} task available in the HEASOFT software package\footnote{https://heasarc.gsfc.nasa.gov/docs/software/heasoft/}.The data was extracted in the 0.5 to 10 keV energy range. For the source extraction, we defined an annular region with inner and outer radii of 60 and 80 arcseconds, respectively. For background extraction, we selected a circular region of radius 70 arcseconds away from the source. The ancillary response file (ARF) was generated using the {\it xrtmkarf} task. We then used the {\it grppha} tool to group the spectra, ensuring that each energy bin contained at least 20 counts. For spectral fitting, we applied an absorbed power-law model. The Galactic hydrogen column density ($N_{H}$) was fixed at 1.34$\times$10$^{20}$$\text{cm}^{-2}$ \citep{2016A&A...594A.116H}.

\section{Analysis and Results}
\label{sec:anr}
\subsection{X-ray polarization}
We analyzed the polarimetric signal from Mrk 421 using {\tt PCUBE} algorithm in the {\tt xpbin} task. We generated three polarization  cubes for the three DUs to extract information including the polarization degree ($\Pi_{X}$) and the polarization angle ($\Psi_{X}$). The combined polarization parameters from the three DUs in the 2$-$3 keV, 3$-$5 keV, 5$-$8 keV and the total 2$-$8 keV bands are given in Table \ref{table-xpol}. The polarization degree $\Pi_{X}$ is found to be 8.5$\pm$0.5\% in the 2$-$8 keV band with $\Psi_{X}$ = 10.6$\pm$1.7 deg. Also, there is an increase in the $\Pi_{X}$ from 2$-$3 keV to 3$-$5 keV. The normalized U/I and Q/I Stokes parameters obtained from the combined cube in different energy bands are shown in Fig. \ref{figure-comp}. We also carried out spectro-polarimetric (model-dependent polarization) analysis. 
We considered Galactic absorption along the line of sight to Mrk 421 by using the \textit{TBabs} model with a weighted average column density ($N_{H}$) from \cite{2016A&A...594A.116H}. The $N_{H}$ value was fixed at 1.34$\times$10$^{20}$$\text{cm}^{-2}$ during the fit. For spectral modeling, we employed a simple power law model (\textit{pow}), and for modeling X-ray polarization, we used the \textit{polconst} model based on Stokes parameters. In XPSEC, the model is expressed as:
\begin{equation} 
\textit{constant} \times \textit{TBabs} \times (\textit{polconst} \times \textit{pow}) 
\end{equation}
Here, \textit{constant} represents the inter-calibration constant for each detector. We achieved statistically acceptable fits with a $\chi^2/dof$ value of 1.3.

The polarization degree $\Pi_{X}$ is found to be 9.5$\pm$1.2\% in the 2$-$8 keV band with $\Psi_{X}$ of 11.1$\pm$3.8 deg. Both methods thus gave consistent results within errors. For the comparison, the derived model-independent and dependent values are given in Table \ref{table-MI_MD}. The position of the measured polarization values in the $\Pi_{X}$ and $\Psi_{X}$ plane in the 2$-$8 keV band for PCUBE and XSPEC analysis are given in Fig. \ref{figure-comp}. Also, the PD change in different energy bands for {\it IXPE} analysis is given in Fig. \ref{figure-comp}.

\begin{table}
\begin{center}
\caption{The measured polarization parameters from model independent and spectro-polarimetry analysis in the 2$-$8 keV band.}\label{table-MI_MD}
 \begin{tabular}{cccccccccccc}
  \hline\noalign{\smallskip}
OBSID  & \multicolumn{2}{c}{Model Independent} & \multicolumn{2}{c}{Spectro-polarimetry}                 \\
  \hline\noalign{\smallskip}
 & $\Pi_X$ (\%) & $\Psi_{X}$(deg) & $\Pi_X$ (\%) & $\Psi_{X}$(deg) \\
\hline\noalign{\smallskip}
02008199  & 8.5$\pm$0.5 & 10.6$\pm$1.7 & 9.5$\pm$1.2 &  11.1$\pm$3.8  \\

  \noalign{\smallskip}\hline
\end{tabular}
\end{center}
\end{table}

\subsection{Optical polarization}

Using AIMPOL, we found optical polarization in the B, V, and R bands. On 06 December 2023, optical polarization degree ($\Pi_O$) was found to be 4.27$\pm$0.32\%, 3.57$\pm$0.31\%, and 3.13$\pm$0.25\% in B, V, and R filters, respectively. In the optical band, we also found an increase in the polarization degree with energy. The results on all optical observations are given in Table \ref{table-oall}. The optical polarization degree and polarization angle for each filter are given in the bottom right panel of Fig.\ref{figure-comp} while the results on all the optical band polarization are given in Fig. \ref{figure-ox}.

\begin{table*}
\centering
\caption{The results of optical polarization observations. Here, N is the number of observations, PD is the polarization degree in
percentage and PA is the position angle in degrees. The quoted polarization PD and PA are the average of the N measurements.}
\label{table-oall}
\begin{tabular}{cccccccccc}
\hline
Date  & \multicolumn{3}{c}{B} & \multicolumn{3}{c}{V} & \multicolumn{3}{c}{R} \\
& N & PD & PA & N & PD & PA & N & PD & PA \\
\hline
02/12/2023 & - &  -& - & - & - & - & 3 & 2.55$\pm$0.24 & 82.9$\pm$0.19 \\

05/12/2023 &3  & 2.99$\pm$0.29 & 124.0$\pm$0.27 & 3 & 1.85$\pm$0.33 & 91.2$\pm$0.23  & 3 & 2.64$\pm$0.28 & 82.7$\pm$0.21 \\

06/12/2023 & 3 & 4.27$\pm$0.32 & 124.0$\pm$0.26 & 3 & 3.57$\pm$0.31 & 127.7$\pm$0.22 & 6 &3.13$\pm$0.25 & 129.1$\pm$0.19 \\

07/12/2023 & 3 & 4.46$\pm$0.34 & 120.5$\pm$0.28 & 3 & 3.56$\pm$0.40 & 80.9$\pm$0.25  & 42 & 3.96$\pm$0.34 & 124.8$\pm$0.23  \\

02/01/2024 & 3 & 6.89$\pm$0.24 & 132.2$\pm$0.42 & 3 & 4.32$\pm$0.21 & 140.4$\pm$0.41 & 3 & 5.70$\pm$0.19 & 125.7$\pm$0.41 \\

03/01/2024 & 3 & 7.51$\pm$0.31 & 142.8$\pm$0.43 & 3 & 5.23$\pm$0.23 & 138.1$\pm$0.42 & 3 & 6.01$\pm$0.22 & 128.4$\pm$0.41 \\
\hline
\end{tabular}
\end{table*}

\section{Discussion}
\label{sec:dis}
Simultaneous optical and X-ray polarization measurements are pivotal in unveiling the underlying emission mechanisms and particle acceleration processes in high-energy sources like blazars. For Mrk 421, a source with multiple X-ray polarization measurements, we conducted a comparison with the values obtained on 06 December 2023 against previous observations \citep{2022ApJ...938L...7D, 2023NatAs...7.1245D, 2023AAS...24133602E,2024AA...681A..12K}. All the available X-ray polarization values of Mrk 421 are given in Table \ref{table-mrk}.
Earlier observations indicate that the degree of X-ray polarization for Mrk 421 ranges from 10\% to 15\%. However, on 06 December 2023, the degree of polarization was found to be 8.5\%, the lowest among all the observations. 
Interestingly, the analysis conducted on 06 December 2022 (ID: 02004401) by \cite{2024AA...681A..12K} revealed a $\Pi_X$ of 14$\pm$1\% and a $\psi_X$ of 107$\pm$3 degrees, with a net exposure time of 74 ks. During this period, the source was in a typical X-ray activity state, exhibiting an X-ray flux of 2.8$\times$10$^{-10}$$erg$ $cm^{-2}$ $s^{-1}$ in the 2$-$8 keV band. In contrast, our analysis of Mrk 421 on 06 December 2023 (ID: 02008199) with a net exposure time of 70 ks, showed a lower $\Pi_X$ of 8.5$\pm$0.5\% and a significantly different $\psi_X$ of 10.6$\pm$1.7 degrees. During this observation, the source exhibited a higher X-ray flux of 3$\times$10$^{-10}$$erg$ $cm^{-2}$ $s^{-1}$ in the 2$-$8 keV band, indicating a state of elevated X-ray activity. This difference in polarization characteristics and X-ray flux suggests a possible change in the
magnetic field structure or emission region within the blazar, reflecting the dynamic nature of high-energy processes in Mrk 421. 
To obtain further insights into the emission processes, we carried out the broadband modeling of its SED using synchrotron and inverse Compton mechanisms.

\begin{table}
\centering
\caption{Summary of X-ray polarization observations available for Mrk 421 till December 2023. The date of observation and the measured degree of polarization is given.}
\label{table-mrk}
\begin{tabular}{c c c }
			\hline
			  Obs. date & $\Pi_{X} (\%$) &  Reference\\
			\hline 
			 2022-05-04 & 15$\pm$2  & \cite{2022ApJ...938L...7D} \\
           2022-06-04 & 10$\pm$1 &  \cite{2023NatAs...7.1245D} \\
           2022-06-07 & 10$\pm$1 & \cite{2023NatAs...7.1245D} \\
           2022-12-06 & 14$\pm$1 & \cite{2024AA...681A..12K} \\
           2023-12-06 & 8.5$\pm$0.5 & This work \\
            \hline 
		\end{tabular}
\end{table}

\begin{figure*}
 \centering
\includegraphics[scale=1]{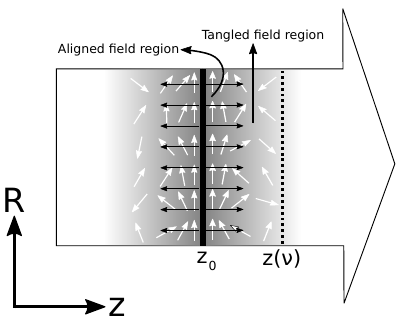}
\includegraphics[scale=0.25]{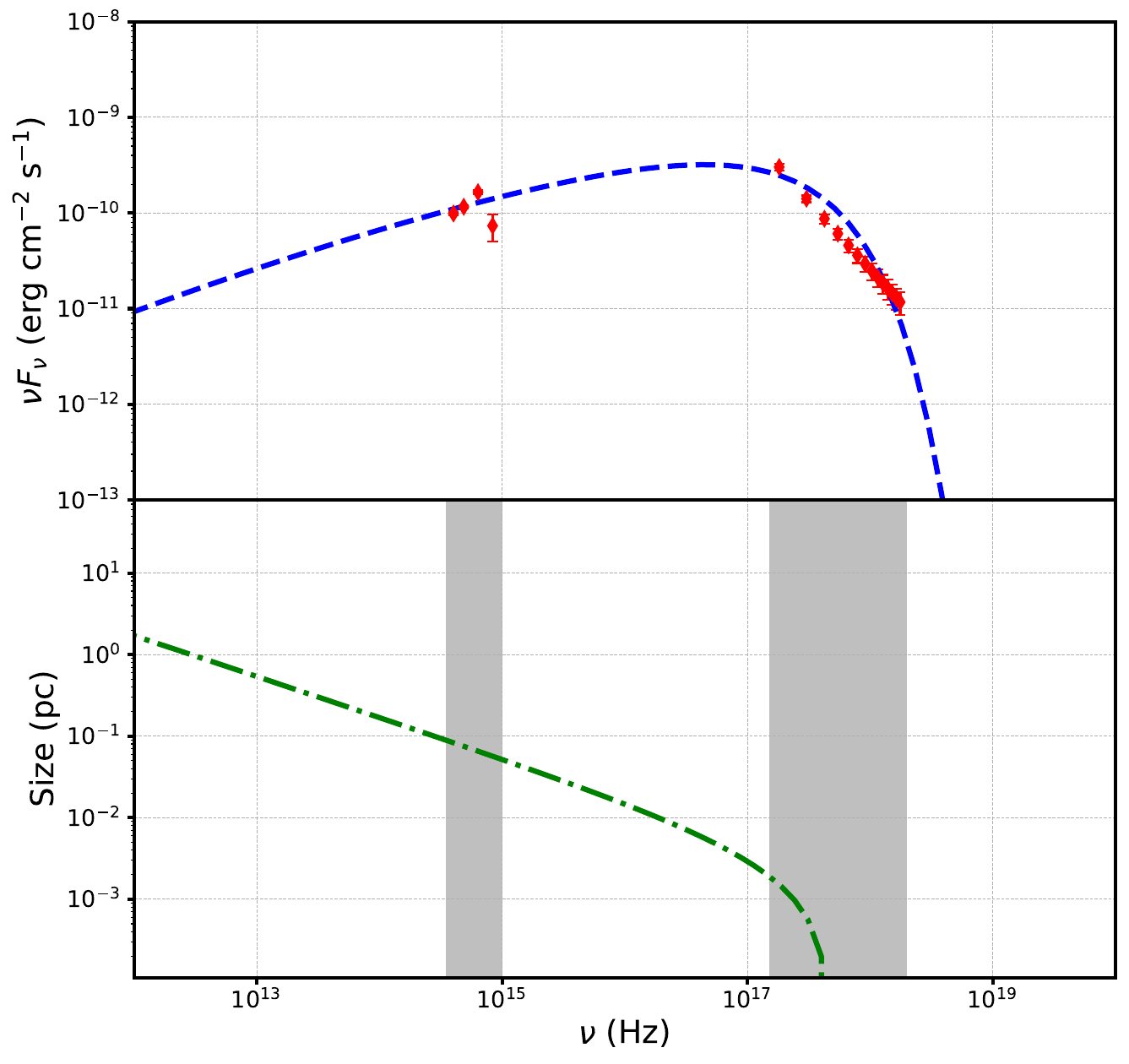}
\caption{ The left panel illustrates the emission regions characterized by different magnetic field configurations: an aligned magnetic field at position $z_{0}$, where X-ray emission originates, and a tangled magnetic field at position $z_{\nu}$, which is the source of optical emission. The right panel consists of two subplots: the top subplot shows the model fit (blue dotted line) to the observed optical and X-ray emissions, while the bottom subplot indicates the spatial extension of their respective emission regions (green dotted line).}
\label{figure-model}
\end{figure*}

Our model assumes the emission region to be located around a shock front. The electrons are accelerated in the vicinity of the shock front and advected into the jet medium while losing their energy through radiative processes. Due to shock compression, the magnetic field  lines close to the shock front will be more ordered and transverse to the flow direction, whereas they may be tangled at regions farther away. Hence, the synchrotron emission happening in the vicinity of the shock front will be more polarized than the farther regions.
Since the X-ray emission involves high energy electrons, the emission will be confined within the vicinity of the shock and will be more polarized. On the other hand, optical/UV emitting electrons demand less energy and can advect to farther regions resulting in less polarized emission. 
 This scenario accounts for the difference in polarization measurements observed in optical and X-ray energies. A schematic of the model considered here is shown in the left panel of Fig. \ref{figure-model}. 

To model the total synchrotron emission arising from this region, we denote the location of the shock front as $z_0$ and the accelerated electron distribution as
\begin{align}
	n(\gamma, z_0) &= k \gamma^{-p} \quad {\rm for} \quad \gamma<\gamma_{\rm max}
\end{align}
where $p$ is the particle index.
Due to synchrotron loss, the distribution at a distance $z$ from $z_0$ will be  
\begin{align}
n(\gamma,z) & =\frac{\bar{P}(\Gamma_z)}{\bar{P}(\gamma)} n(\Gamma_z,z_0)
\end{align} 
where, $\Gamma_z$ is the Lorentz factor of the electron at $z_0$ which reduces to $\gamma$ at $z$ and $\bar{P}$ is
the spatial energy loss. For a constant advection, we can express
\begin{align}
	\bar{P}(\gamma) &= \frac{d \gamma}{d z} \nonumber \\
	&=\left(\frac{d \gamma}{d t}\right)\big/\left(\frac{d z}{d t}\right) \nonumber \\
	&=-\xi \gamma^2 \label{eq:ploss}
\end{align}
and hence,
\begin{align}
n(\gamma,z) =\frac{\Gamma_z^2}{r^2} n(\Gamma_z,z_0)
\end{align} 
Using Equation \ref{eq:ploss}, $\Gamma_z$ can be expressed as
\begin{align}
& \int_{z_0}^z d z=-\int_{\Gamma_z}^{\gamma} \frac{d x}{\xi x^2} 
\end{align}
and we get
\begin{align}
& \Gamma_x=\left[\frac{1}{\gamma}-\xi (z-z_0)\right]^{-1}
\end{align}
Similarly, the extension of the region encompassing the electrons of Lorentz factor $\gamma$ can be 
found as
\begin{align}\label{eq:10}
        z_{\text{max}} (\gamma)= z_0 + \frac{1}{\xi} \left[\frac{1}{\gamma} - \frac{1}{\gamma_{\text{max}}}\right]
\end{align}
The total number of electrons with Lorentz factor $\gamma$ will then be
\begin{align}\label{eq:ntot}
	N_{\text {tot }}(\gamma) & =\pi R^2 \int_{z_0}^{z_{\rm max}(\gamma)} n\left(\Gamma_x, z_0\right) dx 
\end{align}
where, the jet is assumed to be cylindrical with radius $R$. The synchrotron flux received on earth is then 
obtained by convolving Eq.\ref{eq:ntot} with the single particle emissivity \citep{1986rpa..book.....R}. Due to the relativistic motion of the jet the observer will see an enhanced emission modulated by the Doppler factor ($\delta$) and the observed flux after accounting for the cosmological effects will be \citep{1995ApJ...446L..63D}.,
\begin{equation}
    F_{\text {obs }}\left(\nu_{\text {obs }}\right)=\frac{\delta_D^3(1+z)}{d_L^2} V j_{\text {syn }}\left(\frac{1+z}{\delta_D} \nu_{\text {obs }}\right) \quad \mathrm{erg} / \mathrm{cm}^2 / \mathrm{s} / \mathrm{Hz}
\end{equation}
where, $z$ is the redshift of the source, $d_L$ is the luminosity distance, $V$ is the volume of the emission region and $j_{\mathrm{syn}}$ is the emissivity due to synchrotron process.

The model is applied on the simultaneous SED obtained on 06 December, 2023. The X-ray data was reduced from the \textit{Swift}-XRT observation and the optical data was obtained from the observation
using HCT. The fluxes in u$^\prime$, g$^{\prime}$, r$^{\prime}$, and i$^{\prime}$ were used to construct the SED.
The observed optical--X-ray flux points are shown in Fig. \ref{figure-model} (right panel) along with the model curve. 
The model can reproduce the observed SED well and the main parameters are given in Table \ref{table-SED}. 
\begin{table}
\centering
\caption{The model parameters for the SED shown in figure for the epoch of 06 December 2023. The advection  velocity is assumed to be c.}
\label{table-SED}
\begin{tabular}{c c }
			\hline
			  Parameter & Values \\
			\hline 
	        R (cm) & 1.2$\times$10$^{15}$ \\
            B (G) & 0.1  \\
            $\delta$ & 19   \\
            p & 1 \\
            k (\#/$cm^{3}$) & 1   \\
            \hline 
		\end{tabular}
\end{table}
To obtain insight into the observed difference in the optical and X-ray polarization measurements, we study the extension of the regions using the model parameters. The Lorentz factor of the electrons responsible of the observed emission at $v_{\text {obs }}$ will be \citep{1986rpa..book.....R},
\begin{equation}
    v_{\text {obs }}=\frac{\delta}{1+z} \gamma^2 \frac{e B}{2 \pi m c}
\end{equation}
where $B$ is the magnetic field.
Using the parameters provided in Table \ref{table-SED}, we find the electron Lorentz factors responsible for optical and X-ray emission are $\gamma_o$ =8801.39 and $\gamma_X$ =582156.99. The extension of the emission region can then be estimated using
Eq.\ref{eq:10} as $z_o$  = 8.56$\times$10$^{-2}$ pc
and $z_X$ = 1.29$\times$10$^{-3}$ pc. We find the optical and X-ray emitting regions differ by $\sim$0.1 pc. If we attribute the decrease in optical polarization as a result of tangling of the field lines at regions farther from the shock front, then the disorderedness in the field lines happens typically at a length scale of $\sim$0.1 pc.

\section{Summary}
\label{sec:sum}
Mrk 421 is one of the sources observed multiple
times by {\it IXPE}. The new X-ray polarization observation of this source along with simultaneous optical polarization were studied in this work. From our analysis, we arrived at the following conclusions.

\begin{enumerate}
\item  We found significant X-ray polarization value of $\Pi_X$ = 8.5$\pm$0.05\% and $\psi_X$ = 10.6$\pm$1.7 degree in the 2$-$8 keV band from model independent analysis.
This is lower than that observed from the source at earlier epochs of observations.
\item From model-dependent analysis we found  $\Pi_X$ = 9.5$\pm$1.2\% and $\psi_X$ = 11.1$\pm$3.8 degree. This is in agreement with the values obtained from model independent analysis.


\item In the optical band we found the polarization degree to range from 3.13\% to 4.27\% for different filters and this value is found to increase with energies. Also, $\Pi_X$ is found to be larger compared to optical wavelengths. Among X-rays too, we found $\Pi_X$ to increase with energy between 2$-$3 keV and 3$-$5 keV, but drops again in the 5$-$8 keV with a large error that makes the value in the 5$-$8 keV band consistent with the values both at the 2$-$3 keV and 3$-$5 keV bands.  Not considering the polarization degree in the 5$-$8 keV band due to the large error in the measurement, we found an energy dependent polarization from simultaneous observations in optical and X-ray energies. This is similar to that seen in other blazars such as Mrk 501 \citep{2024arXiv240211949H} and 1ES 0229+200 \citep{2023ApJ...959...61E}.

\item Our observations suggest a spatial gradient in the maximum available particle energy within the emission region. This variation in particle energy could potentially explain the observed temporal changes in polarization. This finding aligns with observations of HSP blazars, where studies have shown higher-energy emission exhibiting larger-amplitude polarization variations compared to lower-energy bands.

\item The lower optical polarization compared to X-ray shows that electrons are accelerated in shock front, the high energy X-rays originate close to the shock front with the optical originating at much larger distances down the jet. SED modeling assuming such an scenario suggests the disordereness in the magnetic field lines happens typically on a
length scale of $\sim$0.1 pc from the shock (main acceleration zone).


\end{enumerate}

\begin{acknowledgments}
We thank the anonymous referee for his/her valuable feedback, which has significantly enhanced the quality of this manuscript.
The \textit{Imaging X-ray Polarimetry Explorer} (\textit{IXPE}) is a joint US and Italian mission. The US contribution is supported by the National Aeronautics and Space Administration (NASA) and led and managed by its Marshall Space Flight  Center (MSFC), with industry partner Ball Aerospace (contract NNM15AA18C). The Italian contribution is supported by the Italian Space Agency (Agenzia Spaziale Italiana, ASI) through contract ASI-OHBI-2017-12-I.0, agreements ASI-INAF-2017-12-H0 and ASI-INFN-2017.13-H0, and its Space Science Data Center (SSDC) with agreements ASI- INAF-2022-14- HH.0 and ASI-INFN 2021-43-HH.0, and by the Istituto Nazionale di Astrofisica (INAF) and the Istituto Nazionale di Fisica Nucleare (INFN) in Italy. This research used data products provided by the {\it IXPE} Team (MSFC, SSDC, INAF, and INFN) and distributed with additional software tools by the High-Energy Astrophysics Science Archive Research Center (HEASARC), at NASA Goddard Space Flight Center (GSFC). We acknowledge the use of the Himalayan Chandra Telescope (HCT) at the Indian Astronomical Observatory, Hanle, and the ARIES Imaging Polarimeter (AIMPOL) at the Aryabhatta Research Institute of Observational Sciences (ARIES), Nainital, India, for obtaining the data presented in this work. We extend our gratitude to the staff and researchers at the Indian Institute of Astrophysics (IIA) and ARIES for their support and maintenance of these facilities.
Athira M Bharathan expresses gratitude to the Department of Science and Technology (DST) for the INSPIRE Fellowship (IF200255). AKM acknowledges the support from the European Research Council (ERC) under the European Union’s Horizon 2020 research and innovation programme (grant agreement No. [951549]).

\end{acknowledgments}

\bibliography{mrk_main}{}

\begin{thebibliography}{}
\expandafter\ifx\csname natexlab\endcsname\relax\def\natexlab#1{#1}\fi
\providecommand{\url}[1]{\href{#1}{#1}}
\providecommand{\dodoi}[1]{doi:~\href{http://doi.org/#1}{\nolinkurl{#1}}}
\providecommand{\doeprint}[1]{\href{http://ascl.net/#1}{\nolinkurl{http://ascl.net/#1}}}
\providecommand{\doarXiv}[1]{\href{https://arxiv.org/abs/#1}{\nolinkurl{https://arxiv.org/abs/#1}}}

\bibitem[{{Abdo} {et~al.}(2010){Abdo}, {Ackermann}, {Agudo}, {Ajello}, {Aller}, {Aller}, {Angelakis}, {Arkharov}, {Axelsson}, {Bach}, {Baldini}, {Ballet}, {Barbiellini}, {Bastieri}, {Baughman}, {Bechtol}, {Bellazzini}, {Benitez}, {Berdyugin}, {Berenji}, {Blandford}, {Bloom}, {Boettcher}, {Bonamente}, {Borgland}, {Bregeon}, {Brez}, {Brigida}, {Bruel}, {Burnett}, {Burrows}, {Buson}, {Caliandro}, {Calzoletti}, {Cameron}, {Capalbi}, {Caraveo}, {Carosati}, {Casandjian}, {Cavazzuti}, {Cecchi}, {{\c{C}}elik}, {Charles}, {Chaty}, {Chekhtman}, {Chen}, {Chiang}, {Chincarini}, {Ciprini}, {Claus}, {Cohen-Tanugi}, {Colafrancesco}, {Cominsky}, {Conrad}, {Costamante}, {Cutini}, {D'ammando}, {Deitrick}, {D'Elia}, {Dermer}, {de Angelis}, {de Palma}, {Digel}, {Donnarumma}, {Silva}, {Drell}, {Dubois}, {Dultzin}, {Dumora}, {Falcone}, {Farnier}, {Favuzzi}, {Fegan}, {Focke}, {Forn{\'e}}, {Fortin}, {Frailis}, {Fuhrmann}, {Fukazawa}, {Funk}, {Fusco}, {G{\'o}mez}, {Gargano}, {Gasparrini}, {Gehrels}, {Germani}, {Giebels}, {Giglietto},
  {Giommi}, {Giordano}, {Giuliani}, {Glanzman}, {Godfrey}, {Grenier}, {Gronwall}, {Grove}, {Guillemot}, {Guiriec}, {Gurwell}, {Hadasch}, {Hanabata}, {Harding}, {Hayashida}, {Hays}, {Healey}, {Heidt}, {Hiriart}, {Horan}, {Hoversten}, {Hughes}, {Itoh}, {Jackson}, {J{\'o}hannesson}, {Johnson}, {Johnson}, {Jorstad}, {Kadler}, {Kamae}, {Katagiri}, {Kataoka}, {Kawai}, {Kennea}, {Kerr}, {Kimeridze}, {Kn{\"o}dlseder}, {Kocian}, {Kopatskaya}, {Koptelova}, {Konstantinova}, {Kovalev}, {Kovalev}, {Kurtanidze}, {Kuss}, {Lande}, {Larionov}, {Latronico}, {Leto}, {Lindfors}, {Longo}, {Loparco}, {Lott}, {Lovellette}, {Lubrano}, {Madejski}, {Makeev}, {Marchegiani}, {Marscher}, {Marshall}, {Max-Moerbeck}, {Mazziotta}, {McConville}, {McEnery}, {Meurer}, {Michelson}, {Mitthumsiri}, {Mizuno}, {Moiseev}, {Monte}, {Monzani}, {Morselli}, {Moskalenko}, {Murgia}, {Nestoras}, {Nilsson}, {Nizhelsky}, {Nolan}, {Norris}, {Nuss}, {Ohsugi}, {Ojha}, {Omodei}, {Orlando}, {Ormes}, {Osborne}, {Ozaki}, {Pacciani}, {Padovani}, {Pagani}, {Page},
  {Paneque}, {Panetta}, {Parent}, {Pasanen}, {Pavlidou}, {Pelassa}, {Pepe}, {Perri}, {Pesce-Rollins}, {Piranomonte}, {Piron}, {Pittori}, {Porter}, {Puccetti}, {Rahoui}, {Rain{\`o}}, {Raiteri}, {Rando}, {Razzano}, {Reimer}, {Reimer}, {Reposeur}, {Richards}, {Ritz}, {Rochester}, {Rodriguez}, {Romani}, {Ros}, {Roth}, {Roustazadeh}, {Ryde}, {Sadrozinski}, {Sadun}, {Sanchez}, {Sander}, {Saz Parkinson}, {Scargle}, {Sellerholm}, {Sgr{\`o}}, {Shaw}, {Sigua}, {Siskind}, {Smith}, {Smith}, {Spandre}, {Spinelli}, {Starck}, {Stevenson}, {Stratta}, {Strickman}, {Suson}, {Tajima}, {Takahashi}, {Takahashi}, {Takalo}, {Tanaka}, {Thayer}, {Thayer}, {Thompson}, {Tibaldo}, {Torres}, {Tosti}, {Tramacere}, {Uchiyama}, {Usher}, {Vasileiou}, {Verrecchia}, {Vilchez}, {Villata}, {Vitale}, {Waite}, {Wang}, {Winer}, {Wood}, {Ylinen}, {Zensus}, {Zhekanis}, \& {Ziegler}}]{2010ApJ...716...30A}
{Abdo}, A.~A., {Ackermann}, M., {Agudo}, I., {et~al.} 2010, \apj, 716, 30, \dodoi{10.1088/0004-637X/716/1/30}

\bibitem[{{Acciari} {et~al.}(2009){Acciari}, {Aliu}, {Aune}, {Beilicke}, {Benbow}, {B{\"o}ttcher}, {Bradbury}, {Buckley}, {Bugaev}, {Butt}, {Cannon}, {Celik}, {Cesarini}, {Chow}, {Ciupik}, {Cogan}, {Colin}, {Cui}, {Dickherber}, {Duke}, {Falcone}, {Fegan}, {Finley}, {Finnegan}, {Fortin}, {Fortson}, {Furniss}, {Gall}, {Gillanders}, {Grube}, {Guenette}, {Gyuk}, {Hanna}, {Holder}, {Horan}, {Hui}, {Humensky}, {Kaaret}, {Karlsson}, {Kertzman}, {Kieda}, {Kildea}, {Konopelko}, {Krawczynski}, {Krennrich}, {Lang}, {LeBohec}, {Maier}, {McCann}, {Millis}, {Moriarty}, {Ong}, {Otte}, {Pandel}, {Perkins}, {Pichel}, {Pohl}, {Quinn}, {Ragan}, {Reyes}, {Reynolds}, {Roache}, {Rose}, {Schroedter}, {Sembroski}, {Smith}, {Steele}, {Swordy}, {Theiling}, {Toner}, {Varlotta}, {Vincent}, {Wakely}, {Ward}, {Weekes}, {Weinstein}, {Weisgarber}, {Williams}, {Wissel}, {Zitzer}, {VERITAS Collaboration}, {de la Calle Perez}, {Ibarra}, {Anderhub}, {Rodriguez}, {Antonelli}, {Antoranz}, {Backes}, {Baixeras}, {Balestra}, {Barrio}, {Bastieri},
  {Becerra Gonz{\'a}lez}, {Becker}, {Bednarek}, {Berger}, {Bernardini}, {Biland}, {Bock}, {Bonnoli}, {Bordas}, {Borla Tridon}, {Bosch-Ramon}, {Bose}, {Braun}, {Bretz}, {Britvitch}, {Camara}, {Carmona}, {Carosi}, {Commichau}, {Contreras}, {Cortina}, {Costado}, {Covino}, {Curtef}, {Dazzi}, {DeAngelis}, {DeCea del Pozo}, {Delgado Mendez}, {Delos Reyes}, {DeLotto}, {DeMaria}, {DeSabata}, {Dominguez}, {Dorner}, {Doro}, {Elsaesser}, {Errando}, {Ferenc}, {Fern{\'a}ndez}, {Firpo}, {Fonseca}, {Font}, {Galante}, {Garc{\'\i}a L{\'o}pez}, {Garczarczyk}, {Gaug}, {Goebel}, {Hadasch}, {Hayashida}, {Herrero}, {Hildebrand}, {H{\"o}hne-M{\"o}nch}, {Hose}, {Hsu}, {Jogler}, {Kranich}, {La Barbera}, {Laille}, {Leonardo}, {Lindfors}, {Lombardi}, {Longo}, {L{\'o}pez}, {Lorenz}, {Majumdar}, {Maneva}, {Mankuzhiyil}, {Mannheim}, {Maraschi}, {Mariotti}, {Mart{\'\i}nez}, {Mazin}, {Meucci}, {Miranda}, {Mirzoyan}, {Miyamoto}, {Mold{\'o}n}, {Moles}, {Moralejo}, {Nieto}, {Nilsson}, {Ninkovic}, {Orito}, {Oya}, {Paoletti}, {Paredes},
  {Pasanen}, {Pascoli}, {Pauss}, {Pegna}, {Perez-Torres}, {Persic}, {Peruzzo}, {Prada}, {Prandini}, {Puchades}, {Reichardt}, {Rhode}, {Rib{\'o}}, {Rico}, {Rissi}, {Robert}, {R{\"u}gamer}, {Saggion}, {Saito}, {Salvati}, {Sanchez-Conde}, {Satalecka}, {Scalzotto}, {Scapin}, {Schweizer}, {Shayduk}, {Shore}, {Sidro}, {Sierpowska-Bartosik}, {Sillanp{\"a}{\"a}}, {Sitarek}, {Sobczynska}, {Spanier}, {Spiro}, {Stamerra}, {Stark}, {Takalo}, {Tavecchio}, {Temnikov}, {Tescaro}, {Teshima}, {Tluczykont}, {Torres}, {Turini}, {Vankov}, {Wagner}, {Zabalza}, {Zandanel}, {Zanin}, {Zapatero}, \& {MAGIC Collaboration}}]{2009ApJ...703..169A}
{Acciari}, V.~A., {Aliu}, E., {Aune}, T., {et~al.} 2009, \apj, 703, 169, \dodoi{10.1088/0004-637X/703/1/169}

\bibitem[{{Andruchow} {et~al.}(2005){Andruchow}, {Romero}, \& {Cellone}}]{2005A&A...442...97A}
{Andruchow}, I., {Romero}, G.~E., \& {Cellone}, S.~A. 2005, \aap, 442, 97, \dodoi{10.1051/0004-6361:20053325}

\bibitem[{{Angel} \& {Stockman}(1980)}]{1980ARA&A..18..321A}
{Angel}, J.~R.~P., \& {Stockman}, H.~S. 1980, \araa, 18, 321, \dodoi{10.1146/annurev.aa.18.090180.001541}

\bibitem[{{Baldini} {et~al.}(2022){Baldini}, {Bucciantini}, {Lalla}, {Ehlert}, {Manfreda}, {Negro}, {Omodei}, {Pesce-Rollins}, {Sgr{\`o}}, \& {Silvestri}}]{2022SoftX..1901194B}
{Baldini}, L., {Bucciantini}, N., {Lalla}, N.~D., {et~al.} 2022, SoftwareX, 19, 101194, \dodoi{10.1016/j.softx.2022.101194}

\bibitem[{{Burrows} {et~al.}(2005){Burrows}, {Hill}, {Nousek}, {Kennea}, {Wells}, {Osborne}, {Abbey}, {Beardmore}, {Mukerjee}, {Short}, {Chincarini}, {Campana}, {Citterio}, {Moretti}, {Pagani}, {Tagliaferri}, {Giommi}, {Capalbi}, {Tamburelli}, {Angelini}, {Cusumano}, {Br{\"a}uninger}, {Burkert}, \& {Hartner}}]{2005SSRv..120..165B}
{Burrows}, D.~N., {Hill}, J.~E., {Nousek}, J.~A., {et~al.} 2005, \ssr, 120, 165, \dodoi{10.1007/s11214-005-5097-2}

\bibitem[{{Dermer}(1995)}]{1995ApJ...446L..63D}
{Dermer}, C.~D. 1995, \apjl, 446, L63, \dodoi{10.1086/187931}

\bibitem[{{Di Gesu} {et~al.}(2022){Di Gesu}, {Donnarumma}, {Tavecchio}, {Agudo}, {Barnounin}, {Cibrario}, {Di Lalla}, {Di Marco}, {Escudero}, {Errando}, {Jorstad}, {Kim}, {Kouch}, {Liodakis}, {Lindfors}, {Madejski}, {Marshall}, {Marscher}, {Middei}, {Muleri}, {Myserlis}, {Negro}, {Omodei}, {Pacciani}, {Paggi}, {Perri}, {Puccetti}, {Antonelli}, {Bachetti}, {Baldini}, {Baumgartner}, {Bellazzini}, {Bianchi}, {Bongiorno}, {Bonino}, {Brez}, {Bucciantini}, {Capitanio}, {Castellano}, {Cavazzuti}, {Ciprini}, {Costa}, {De Rosa}, {Del Monte}, {Doroshenko}, {Dov{\v{c}}iak}, {Ehlert}, {Enoto}, {Evangelista}, {Fabiani}, {Ferrazzoli}, {Garcia}, {Gunji}, {Hayashida}, {Heyl}, {Iwakiri}, {Karas}, {Kitaguchi}, {Kolodziejczak}, {Krawczynski}, {La Monaca}, {Latronico}, {Maldera}, {Manfreda}, {Marin}, {Marinucci}, {Massaro}, {Matt}, {Mitsuishi}, {Mizuno}, {Ng}, {O'Dell}, {Oppedisano}, {Papitto}, {Pavlov}, {Peirson}, {Pesce-Rollins}, {Petrucci}, {Pilia}, {Possenti}, {Poutanen}, {Ramsey}, {Rankin}, {Ratheesh}, {Romani}, {Sgr{\`o}},
  {Slane}, {Soffitta}, {Spandre}, {Tamagawa}, {Taverna}, {Tawara}, {Tennant}, {Thomas}, {Tombesi}, {Trois}, {Tsygankov}, {Turolla}, {Vink}, {Weisskopf}, {Wu}, {Xie}, \& {Zane}}]{2022ApJ...938L...7D}
{Di Gesu}, L., {Donnarumma}, I., {Tavecchio}, F., {et~al.} 2022, \apjl, 938, L7, \dodoi{10.3847/2041-8213/ac913a}

\bibitem[{{Di Gesu} {et~al.}(2023){Di Gesu}, {Marshall}, {Ehlert}, {Kim}, {Donnarumma}, {Tavecchio}, {Liodakis}, {Kiehlmann}, {Agudo}, {Jorstad}, {Muleri}, {Marscher}, {Puccetti}, {Middei}, {Perri}, {Pacciani}, {Negro}, {Romani}, {Di Marco}, {Blinov}, {Bourbah}, {Kontopodis}, {Mandarakas}, {Romanopoulos}, {Skalidis}, {Vervelaki}, {Casadio}, {Escudero}, {Myserlis}, {Gurwell}, {Rao}, {Keating}, {Kouch}, {Lindfors}, {Aceituno}, {Bernardos}, {Bonnoli}, {Casanova}, {Garc{\'\i}a-Comas}, {Ag{\'\i}s-Gonz{\'a}lez}, {Husillos}, {Marchini}, {Sota}, {Imazawa}, {Sasada}, {Fukazawa}, {Kawabata}, {Uemura}, {Mizuno}, {Nakaoka}, {Akitaya}, {Savchenko}, {Vasilyev}, {G{\'o}mez}, {Antonelli}, {Barnouin}, {Bonino}, {Cavazzuti}, {Costamante}, {Chen}, {Cibrario}, {De Rosa}, {Di Pierro}, {Errando}, {Kaaret}, {Karas}, {Krawczynski}, {Lisalda}, {Madejski}, {Malacaria}, {Marin}, {Marinucci}, {Massaro}, {Matt}, {Mitsuishi}, {O'Dell}, {Paggi}, {Peirson}, {Petrucci}, {Ramsey}, {Tennant}, {Wu}, {Bachetti}, {Baldini}, {Baumgartner},
  {Bellazzini}, {Bianchi}, {Bongiorno}, {Brez}, {Bucciantini}, {Capitanio}, {Castellano}, {Ciprini}, {Costa}, {Del Monte}, {Di Lalla}, {Doroshenko}, {Dov{\v{c}}iak}, {Enoto}, {Evangelista}, {Fabiani}, {Ferrazzoli}, {Garcia}, {Gunji}, {Hayashida}, {Heyl}, {Iwakiri}, {Kislat}, {Kitaguchi}, {Kolodziejczak}, {La Monaca}, {Latronico}, {Maldera}, {Manfreda}, {Ng}, {Omodei}, {Oppedisano}, {Papitto}, {Pavlov}, {Pesce-Rollins}, {Pilia}, {Possenti}, {Poutanen}, {Rankin}, {Ratheesh}, {Roberts}, {Sgr{\`o}}, {Slane}, {Soffitta}, {Spandre}, {Swartz}, {Tamagawa}, {Taverna}, {Tawara}, {Thomas}, {Tombesi}, {Trois}, {Tsygankov}, {Turolla}, {Vink}, {Weisskopf}, {Xie}, \& {Zane}}]{2023NatAs...7.1245D}
{Di Gesu}, L., {Marshall}, H.~L., {Ehlert}, S.~R., {et~al.} 2023, Nature Astronomy, 7, 1245, \dodoi{10.1038/s41550-023-02032-7}

\bibitem[{{Ehlert}(2023)}]{2023AAS...24133602E}
{Ehlert}, S. 2023, in American Astronomical Society Meeting Abstracts, Vol.~55, American Astronomical Society Meeting Abstracts, 336.02

\bibitem[{{Ehlert} {et~al.}(2023){Ehlert}, {Liodakis}, {Middei}, {Marscher}, {Tavecchio}, {Agudo}, {Kouch}, {Lindfors}, {Nilsson}, {Myserlis}, {Gurwell}, {Rao}, {Aceituno}, {Bonnoli}, {Casanova}, {Ag{\'\i}s-Gonz{\'a}lez}, {Escudero}, {Husillos}, {Otero Santos}, {Sota}, {Angelakis}, {Kraus}, {Keating}, {Antonelli}, {Bachetti}, {Baldini}, {Baumgartner}, {Bellazzini}, {Bianchi}, {Bongiorno}, {Bonino}, {Brez}, {Bucciantini}, {Capitanio}, {Castellano}, {Cavazzuti}, {Chen}, {Ciprini}, {Costa}, {De Rosa}, {Del Monte}, {Di Gesu}, {Di Lalla}, {Di Marco}, {Donnarumma}, {Doroshenko}, {Dov{\v{c}}iak}, {Enoto}, {Evangelista}, {Fabiani}, {Ferrazzoli}, {Garcia}, {Gunji}, {Hayashida}, {Heyl}, {Iwakiri}, {Jorstad}, {Kaaret}, {Karas}, {Kislat}, {Kitaguchi}, {Kolodziejczak}, {Krawczynski}, {La Monaca}, {Latronico}, {Maldera}, {Manfreda}, {Marin}, {Marinucci}, {Marshall}, {Massaro}, {Matt}, {Mitsuishi}, {Mizuno}, {Muleri}, {Negro}, {Ng}, {O'Dell}, {Omodei}, {Oppedisano}, {Papitto}, {Pavlov}, {Peirson}, {Perri}, {Pesce-Rollins},
  {Petrucci}, {Pilia}, {Possenti}, {Poutanen}, {Puccetti}, {Ramsey}, {Rankin}, {Ratheesh}, {Roberts}, {Romani}, {Sgr{\'o}}, {Slane}, {Soffitta}, {Spandre}, {Swartz}, {Tamagawa}, {Taverna}, {Tawara}, {Tennant}, {Thomas}, {Tombesi}, {Trois}, {Tsygankov}, {Turolla}, {Vink}, {Weisskopf}, {Wu}, {Xie}, \& {Zane}}]{2023ApJ...959...61E}
{Ehlert}, S.~R., {Liodakis}, I., {Middei}, R., {et~al.} 2023, \apj, 959, 61, \dodoi{10.3847/1538-4357/ad05c4}

\bibitem[{{Fossati} {et~al.}(1998){Fossati}, {Maraschi}, {Celotti}, {Comastri}, \& {Ghisellini}}]{1998MNRAS.299..433F}
{Fossati}, G., {Maraschi}, L., {Celotti}, A., {Comastri}, A., \& {Ghisellini}, G. 1998, \mnras, 299, 433, \dodoi{10.1046/j.1365-8711.1998.01828.x}

\bibitem[{{HI4PI Collaboration} {et~al.}(2016){HI4PI Collaboration}, {Ben Bekhti}, {Fl{\"o}er}, {Keller}, {Kerp}, {Lenz}, {Winkel}, {Bailin}, {Calabretta}, {Dedes}, {Ford}, {Gibson}, {Haud}, {Janowiecki}, {Kalberla}, {Lockman}, {McClure-Griffiths}, {Murphy}, {Nakanishi}, {Pisano}, \& {Staveley-Smith}}]{2016A&A...594A.116H}
{HI4PI Collaboration}, {Ben Bekhti}, N., {Fl{\"o}er}, L., {et~al.} 2016, \aap, 594, A116, \dodoi{10.1051/0004-6361/201629178}

\bibitem[{{Hovatta} \& {Lindfors}(2019)}]{2019NewAR..8701541H}
{Hovatta}, T., \& {Lindfors}, E. 2019, \nar, 87, 101541, \dodoi{10.1016/j.newar.2020.101541}

\bibitem[{{Hu} {et~al.}(2024){Hu}, {Yu}, {Zhang}, {Wang}, {Patra}, {Brink}, {Zheng}, {Wang}, {Kong}, {Chen}, {Zhou}, {Cao}, {Lu}, {Zhou}, {Wei}, {Huang}, {Li}, {Lou}, {Mao}, {Liang}, \& {Filippenko}}]{2024arXiv240211949H}
{Hu}, X.-K., {Yu}, Y.-W., {Zhang}, J., {et~al.} 2024, arXiv e-prints, arXiv:2402.11949, \dodoi{10.48550/arXiv.2402.11949}

\bibitem[{{Kim} {et~al.}(2024{\natexlab{a}}){Kim}, {Di Gesu}, {Liodakis}, {Marscher}, {Jorstad}, {Middei}, {Marshall}, {Pacciani}, {Agudo}, {Tavecchio}, {Cibrario}, {Tugliani}, {Bonino}, {Negro}, {Puccetti}, {Tombesi}, {Costa}, {Donnarumma}, {Soffitta}, {Mizuno}, {Fukazawa}, {Kawabata}, {Nakaoka}, {Uemura}, {Imazawa}, {Sasada}, {Akitaya}, {Jos{\`e} Aceituno}, {Bonnoli}, {Casanova}, {Myserlis}, {Sievers}, {Angelakis}, {Kraus}, {Yeon Cheong}, {Jeong}, {Kang}, {Kim}, {Lee}, {Ag{\`\i}s-Gonz{\`a}lez}, {Sota}, {Escudero}, {Gurwell}, {Keating}, {Rao}, {Kouch}, {Lindfors}, {Bourbah}, {Kiehlmann}, {Kontopodis}, {Mandarakas}, {Romanopoulos}, {Skalidis}, {Vervelaki}, {Savchenko}, {Antonelli}, {Bachetti}, {Baldini}, {Baumgartner}, {Bellazzini}, {Bianchi}, {Bongiorno}, {Brez}, {Bucciantini}, {Capitanio}, {Castellano}, {Cavazzuti}, {Chen}, {Ciprini}, {De Rosa}, {Del Monte}, {Di Lalla}, {Di Marco}, {Doroshenko}, {Dov{\v{c}}iak}, {Ehlert}, {Enoto}, {Evangelista}, {Fabiani}, {Ferrazzoli}, {Garcia}, {Gunji}, {Hayashida},
  {Heyl}, {Iwakiri}, {Kaaret}, {Karas}, {Kislat}, {Kitaguchi}, {Kolodziejczak}, {Krawczynski}, {La Monaca}, {Latronico}, {Maldera}, {Manfreda}, {Marin}, {Marinucci}, {Massaro}, {Matt}, {Mitsuishi}, {Muleri}, {Ng}, {O'Dell}, {Omodei}, {Oppedisano}, {Papitto}, {Pavlov}, {Peirson}, {Perri}, {Pesce-Rollins}, {Petrucci}, {Pilia}, {Possenti}, {Poutanen}, {Ramsey}, {Rankin}, {Ratheesh}, {Roberts}, {Romani}, {Sgr{\'o}}, {Slane}, {Spandre}, {Swartz}, {Tamagawa}, {Taverna}, {Tawara}, {Tennant}, {Thomas}, {Trois}, {Tsygankov}, {Turolla}, {Vink}, {Weisskopf}, {Wu}, {Xie}, \& {Zane}}]{2024A&A...681A..12K}
{Kim}, D.~E., {Di Gesu}, L., {Liodakis}, I., {et~al.} 2024{\natexlab{a}}, \aap, 681, A12, \dodoi{10.1051/0004-6361/202347408}

\bibitem[{{Kim} {et~al.}(2024{\natexlab{b}}){Kim}, {Di Gesu}, {Liodakis}, {Marscher}, {Jorstad}, {Middei}, {Marshall}, {Pacciani}, {Agudo}, {Tavecchio}, {Cibrario}, {Tugliani}, {Bonino}, {Negro}, {Puccetti}, {Tombesi}, {Costa}, {Donnarumma}, {Soffitta}, {Mizuno}, {Fukazawa}, {Kawabata}, {Nakaoka}, {Uemura}, {Imazawa}, {Sasada}, {Akitaya}, {Jos{\`e} Aceituno}, {Bonnoli}, {Casanova}, {Myserlis}, {Sievers}, {Angelakis}, {Kraus}, {Yeon Cheong}, {Jeong}, {Kang}, {Kim}, {Lee}, {Ag{\`\i}s-Gonz{\`a}lez}, {Sota}, {Escudero}, {Gurwell}, {Keating}, {Rao}, {Kouch}, {Lindfors}, {Bourbah}, {Kiehlmann}, {Kontopodis}, {Mandarakas}, {Romanopoulos}, {Skalidis}, {Vervelaki}, {Savchenko}, {Antonelli}, {Bachetti}, {Baldini}, {Baumgartner}, {Bellazzini}, {Bianchi}, {Bongiorno}, {Brez}, {Bucciantini}, {Capitanio}, {Castellano}, {Cavazzuti}, {Chen}, {Ciprini}, {De Rosa}, {Del Monte}, {Di Lalla}, {Di Marco}, {Doroshenko}, {Dov{\v{c}}iak}, {Ehlert}, {Enoto}, {Evangelista}, {Fabiani}, {Ferrazzoli}, {Garcia}, {Gunji}, {Hayashida},
  {Heyl}, {Iwakiri}, {Kaaret}, {Karas}, {Kislat}, {Kitaguchi}, {Kolodziejczak}, {Krawczynski}, {La Monaca}, {Latronico}, {Maldera}, {Manfreda}, {Marin}, {Marinucci}, {Massaro}, {Matt}, {Mitsuishi}, {Muleri}, {Ng}, {O'Dell}, {Omodei}, {Oppedisano}, {Papitto}, {Pavlov}, {Peirson}, {Perri}, {Pesce-Rollins}, {Petrucci}, {Pilia}, {Possenti}, {Poutanen}, {Ramsey}, {Rankin}, {Ratheesh}, {Roberts}, {Romani}, {Sgr{\'o}}, {Slane}, {Spandre}, {Swartz}, {Tamagawa}, {Taverna}, {Tawara}, {Tennant}, {Thomas}, {Trois}, {Tsygankov}, {Turolla}, {Vink}, {Weisskopf}, {Wu}, {Xie}, \& {Zane}}]{2024AA...681A..12K}
---. 2024{\natexlab{b}}, \aap, 681, A12, \dodoi{10.1051/0004-6361/202347408}

\bibitem[{{Lynden-Bell}(1969)}]{1969Natur.223..690L}
{Lynden-Bell}, D. 1969, \nat, 223, 690, \dodoi{10.1038/223690a0}

\bibitem[{{Pandey} {et~al.}(2023){Pandey}, {Singh}, {Yadav}, {Nanjappa}, {Pant}, {Kumar}, \& {Sahu}}]{2023JAI....1240008P}
{Pandey}, J.~C., {Singh}, S., {Yadav}, R.~K.~S., {et~al.} 2023, Journal of Astronomical Instrumentation, 12, 2240008, \dodoi{10.1142/S2251171722400086}

\bibitem[{{Ramaprakash} {et~al.}(1998){Ramaprakash}, {Gupta}, {Sen}, \& {Tandon}}]{1998A&AS..128..369R}
{Ramaprakash}, A.~N., {Gupta}, R., {Sen}, A.~K., \& {Tandon}, S.~N. 1998, \aaps, 128, 369, \dodoi{10.1051/aas:1998148}

\bibitem[{{Rautela} {et~al.}(2004){Rautela}, {Joshi}, \& {Pandey}}]{2004BASI...32..159R}
{Rautela}, B.~S., {Joshi}, G.~C., \& {Pandey}, J.~C. 2004, Bulletin of the Astronomical Society of India, 32, 159

\bibitem[{{Rybicki} \& {Lightman}(1986)}]{1986rpa..book.....R}
{Rybicki}, G.~B., \& {Lightman}, A.~P. 1986, {Radiative Processes in Astrophysics}

\bibitem[{{Schmidt} {et~al.}(1992){Schmidt}, {Elston}, \& {Lupie}}]{1992AJ....104.1563S}
{Schmidt}, G.~D., {Elston}, R., \& {Lupie}, O.~L. 1992, \aj, 104, 1563, \dodoi{10.1086/116341}

\bibitem[{{Shakura} \& {Sunyaev}(1973)}]{1973A&A....24..337S}
{Shakura}, N.~I., \& {Sunyaev}, R.~A. 1973, \aap, 24, 337

\bibitem[{{Tody}(1986)}]{1986SPIE..627..733T}
{Tody}, D. 1986, in Society of Photo-Optical Instrumentation Engineers (SPIE) Conference Series, Vol. 627, Instrumentation in astronomy VI, ed. D.~L. {Crawford}, 733, \dodoi{10.1117/12.968154}

\bibitem[{{Tody}(1993)}]{1993ASPC...52..173T}
{Tody}, D. 1993, in Astronomical Society of the Pacific Conference Series, Vol.~52, Astronomical Data Analysis Software and Systems II, ed. R.~J. {Hanisch}, R.~J.~V. {Brissenden}, \& J.~{Barnes}, 173

\bibitem[{{Ulrich} {et~al.}(1997){Ulrich}, {Maraschi}, \& {Urry}}]{1997ARA&A..35..445U}
{Ulrich}, M.-H., {Maraschi}, L., \& {Urry}, C.~M. 1997, \araa, 35, 445, \dodoi{10.1146/annurev.astro.35.1.445}

\bibitem[{{Urry} \& {Padovani}(1995)}]{1995PASP..107..803U}
{Urry}, C.~M., \& {Padovani}, P. 1995, \pasp, 107, 803, \dodoi{10.1086/133630}

\bibitem[{{Wagner} \& {Witzel}(1995)}]{1995ARA&A..33..163W}
{Wagner}, S.~J., \& {Witzel}, A. 1995, \araa, 33, 163, \dodoi{10.1146/annurev.aa.33.090195.001115}

\bibitem[{{Weisskopf} {et~al.}(2022){Weisskopf}, {Soffitta}, {Baldini}, {Ramsey}, {O'Dell}, {Romani}, {Matt}, {Deininger}, {Baumgartner}, {Bellazzini}, {Costa}, {Kolodziejczak}, {Latronico}, {Marshall}, {Muleri}, {Bongiorno}, {Tennant}, {Bucciantini}, {Dovciak}, {Marin}, {Marscher}, {Poutanen}, {Slane}, {Turolla}, {Kalinowski}, {Di Marco}, {Fabiani}, {Minuti}, {La Monaca}, {Pinchera}, {Rankin}, {Sgro'}, {Trois}, {Xie}, {Alexander}, {Allen}, {Amici}, {Andersen}, {Antonelli}, {Antoniak}, {Attin{\`a}}, {Barbanera}, {Bachetti}, {Baggett}, {Bladt}, {Brez}, {Bonino}, {Boree}, {Borotto}, {Breeding}, {Brienza}, {Bygott}, {Caporale}, {Cardelli}, {Carpentiero}, {Castellano}, {Castronuovo}, {Cavalli}, {Cavazzuti}, {Ceccanti}, {Centrone}, {Citraro}, {D'Amico}, {D'Alba}, {Di Gesu}, {Del Monte}, {Dietz}, {Di Lalla}, {Persio}, {Dolan}, {Donnarumma}, {Evangelista}, {Ferrant}, {Ferrazzoli}, {Ferrie}, {Footdale}, {Forsyth}, {Foster}, {Garelick}, {Gunji}, {Gurnee}, {Head}, {Hibbard}, {Johnson}, {Kelly}, {Kilaru}, {Lefevre},
  {Roy}, {Loffredo}, {Lorenzi}, {Lucchesi}, {Maddox}, {Magazzu}, {Maldera}, {Manfreda}, {Mangraviti}, {Marengo}, {Marrocchesi}, {Massaro}, {Mauger}, {McCracken}, {McEachen}, {Mize}, {Mereu}, {Mitchell}, {Mitsuishi}, {Morbidini}, {Mosti}, {Nasimi}, {Negri}, {Negro}, {Nguyen}, {Nitschke}, {Nuti}, {Onizuka}, {Oppedisano}, {Orsini}, {Osborne}, {Pacheco}, {Paggi}, {Painter}, {Pavelitz}, {Pentz}, {Piazzolla}, {Perri}, {Pesce-Rollins}, {Peterson}, {Pilia}, {Profeti}, {Puccetti}, {Ranganathan}, {Ratheesh}, {Reedy}, {Root}, {Rubini}, {Ruswick}, {Sanchez}, {Sarra}, {Santoli}, {Scalise}, {Sciortino}, {Schroeder}, {Seek}, {Sosdian}, {Spandre}, {Speegle}, {Tamagawa}, {Tardiola}, {Tobia}, {Thomas}, {Valerie}, {Vimercati}, {Walden}, {Weddendorf}, {Wedmore}, {Welch}, {Zanetti}, \& {Zanetti}}]{2022JATIS...8b6002W}
{Weisskopf}, M.~C., {Soffitta}, P., {Baldini}, L., {et~al.} 2022, Journal of Astronomical Telescopes, Instruments, and Systems, 8, 026002, \dodoi{10.1117/1.JATIS.8.2.026002}

\end{thebibliography}
\bibliographystyle{aasjournal}

\end{document}